\documentclass[runningheads]{llncs}
\usepackage{floatrow}
\usepackage{amsfonts, amssymb}
\usepackage{mathrsfs}
\usepackage{graphicx}
\usepackage{subfig}
\usepackage{booktabs}
\usepackage{dblfloatfix}

\begin{document}

%
\title{Unsupervised Domain Adaptation via Disentangled Representations: Application to Cross-Modality Liver Segmentation}
\titlerunning{Unsupervised Domain Adaptation via Disentangled Representations}
%
\author{Junlin Yang\inst{1} \and
Nicha C. Dvornek\inst{3} \and
Fan Zhang\inst{2} \and
Julius Chapiro\inst{3} \and
MingDe Lin\inst{3} \and
James S. Duncan\inst{1,2,3,4}}


\authorrunning{J. Yang et al.}
%
\institute{Department of Biomedical Engineering, Yale University, New Haven, CT, USA \\\email{junlin.yang@yale.edu} \and
Department of Electrical Engineering, Yale University, New Haven, CT, USA \and
Department of Radiology \& Biomedical Imaging, Yale School of Medicine, New Haven, CT, USA \and Department of Statistics \& Data Science, Yale University, New Haven, CT, USA}

\maketitle              
\begin{abstract}
A deep learning model trained on some labeled data from a certain source domain generally performs poorly on data from different target domains due to domain shifts. Unsupervised domain adaptation methods address this problem by alleviating the domain shift between the labeled source data and the unlabeled target data. In this work, we achieve cross-modality domain adaptation, i.e. between CT and MRI images, via disentangled representations. Compared to learning a one-to-one mapping as the state-of-art CycleGAN, our model recovers a many-to-many mapping between domains to capture the complex cross-domain relations. It preserves semantic feature-level information by finding a shared content space instead of a direct pixelwise style transfer. Domain adaptation is achieved in two steps. First, images from each domain are embedded into two spaces, a shared domain-invariant content space and a domain-specific style space. Next, the representation in the content space is extracted to perform a task. We validated our method on a cross-modality liver segmentation task, to train a liver segmentation model on CT images that also performs well on MRI. Our method achieved Dice Similarity Coefficient (DSC) of 0.81, outperforming a CycleGAN-based method of 0.72. Moreover, our model achieved good generalization to joint-domain learning, in which unpaired data from different modalities are jointly learned to improve the segmentation performance on each individual modality. Lastly, under a multi-modal target domain with significant diversity, our approach exhibited the potential for diverse image generation and remained effective with DSC of 0.74 on multi-phasic MRI while the CycleGAN-based method performed poorly with a DSC of only 0.52.\let\thefootnote\relax\footnote{This work was supported by NIH Grant 5R01 CA206180}

\end{abstract}
\section{Introduction}

Deep neural networks have been very successful in a variety of computer vision tasks, including medical image analysis. The majority of neural networks conduct training and evaluation on images from the same distribution. However, real-world applications usually face varying visual domains. The distribution differences between training and test data, i.e. domain shifts, can lead to significant performance degradation. Data collection and manual annotation for every new task and domain are time-consuming and expensive, especially for medical imaging, where data are limited and are collected from different scanners, protocols, sites, and modalities. To solve this problem, domain adaptation algorithms look to build a model from a source data distribution that performs well on a different but related target data distribution \cite{wang2018deep}. In the context of medical image analysis, most prior studies on domain adaptation focus on aligning distributions of data from different scan protocols, scanners, and sites \cite{kamnitsas2017unsupervised,perone2019promises,valindria2018domain}. Related literature is relatively limited when it comes to different modalities.

In clinical practice, various imaging modalities may have valuable and complementary roles. For example, as a fast, less expensive, robust, and readily available modality, computed tomography (CT) plays a key role in the routine clinical examination of hepatocellular carcinoma (HCC), but has the disadvantages of radiation exposure and relatively low soft-tissue contrast. Magnetic resonance imaging (MRI) provides higher soft-tissue contrast for lesion detection and characterization but is more expensive, time-consuming, less robust, and more prone to artifacts. In practice, both multiphase contrast-enhanced MRI and CT may be used in the diagnosis and follow-up after treatment of HCC, and often require the same image analysis tasks, such as liver segmentation. While MRI acquisitions include more complex quantitative information than CT useful for liver segmentation, they are often less available clinically than CT images \cite{oliva2004liver}. Thus, it would be helpful if we could learn a liver segmentation model from the more accessible CT data that also performs well on MRI images. 

\begin{figure}[!b]
\begin{center}
\includegraphics[scale=0.33]{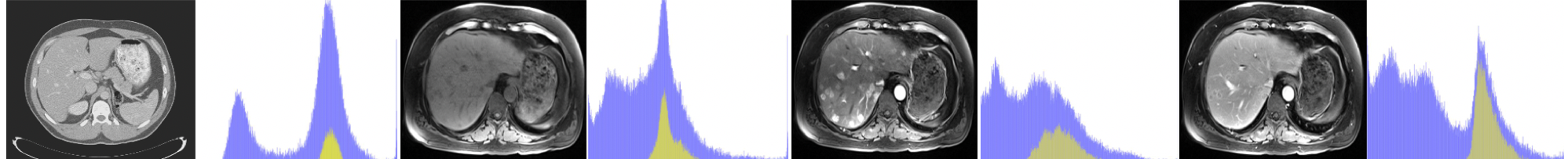}
\end{center}
\caption{Images and histograms of liver (yellow) and whole image (blue). From left to right: CT, multiphasic MRI sequence at three time points (pre-contrast, 20 seconds post-contrast i.e. arterial phase, 70 seconds post-contrast i.e. portal venous phase)}  \label{fig1}
\end{figure}

Given the significant domain shift, cross-modality domain adaptation is quite difficult (see Fig.~\ref{fig1}). One promising approach utilizes CycleGAN, a pixel-wise style transfer model, for cross-modality domain adaptation in a segmentation task \cite{jiang2018tumor}. Compared to feature-based domain adaptation, it does not necessarily maintain the semantic feature-level information. More importantly, the cycle-consistency loss implies a one-to-one mapping between source domain and target domain and leads to lack of translated output diversity, generating very similar images. It thus fails to represent the complex real-world data distribution in the target domain and likely degrades the performance of segmentation or other follow-up analysis \cite{lee2018diverse}.

Our goal is to achieve domain adaptation between CT and MRI while maintaining the complex relationship between the two domains. Our model assumes the mapping to be many-to-many and learns it by disentangling the representation into content and style. A shared latent space is assumed to be found for both domains that preserves the semantic content information. Our main contributions are listed as follows. This is the first to achieve unsupervised domain adaptation for segmentation via disentangled representations in the field of medical imaging. Our model decomposes images across domains into a domain-invariant content space, which preserves the anatomical information, and a domain-specific style space, which represents modality information.  We validated the superior performance of our model on a liver segmentation task with cross-modality domain adaptation and compared it to the state-of-art CycleGAN. We also demonstrated the generalizability of our model to joint-domain learning and robust adaptation to a multi-modal target domain with large variety. 

\section{Methodology}
\subsection{Assumptions}
Let $x_1\in\mathcal{X}_1$ and $x_2\in\mathcal{X}_2$ be images from two domains, which differ in visual appearance but share common semantic content. We assume there exists a mapping, potentially many-to-many instead of deterministic one-to-one, between $\mathcal{X}_1$ and $\mathcal{X}_2$. Each image $x_i$ from $\mathcal{X}_i$ can be embedded into and generated from a shared semantic content space $c\in\mathcal{C}$ that is domain-invariant and a style code $s_i\in\mathcal{S}_i$ that is domain-specific $(i=1,2)$ \cite{huang2018multimodal}. Specifically, MRI and CT of the abdomen from HCC patients can be considered as images from different domains $\mathcal{X}_1$ and $\mathcal{X}_2$, since they exhibit quite different visual appearance with the same anatomical structure shared behind them. Therefore, a shared domain-invariant space that preserves the anatomical information and a domain-specific style code for each modality can be found to recover the underlying mapping between MRI and CT. Due to the many-to-many assumption, the relatively complex underlying distribution of the target domain can be recovered. 

\subsection{Model}

\begin{figure}[!b]
\begin{center}
\includegraphics[scale=0.13]{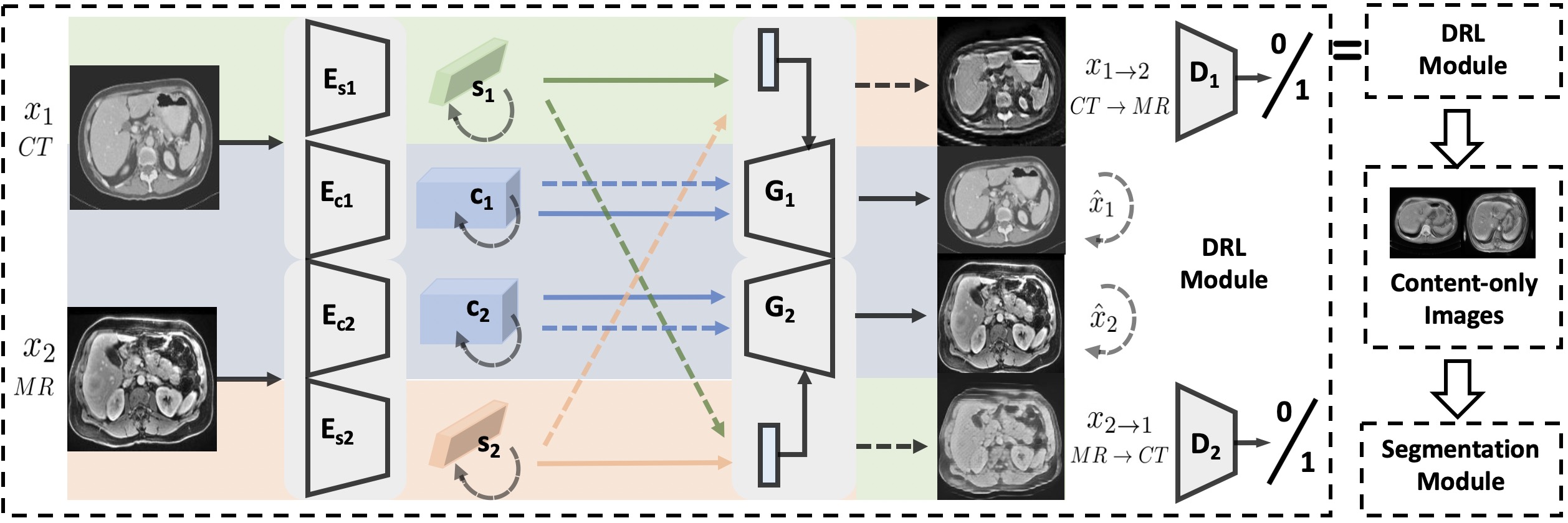}
\end{center}
\caption{Left: Framework for Disentangled Representation Learning Module. Solid line: in-domain reconstruction, Dotted line: cross-domain translation. Right: Pipeline of Domain Adaptation via Disentangled Representations (DADR)} \label{fig2}
\end{figure}

Our Domain Adaptation via Disentangled Representations (DADR) pipeline consists of two modules: Disentangled Representation Learning Module (DRLModule) and Segmentation Module (SegModule) (see Fig.~\ref{fig2}). Of note, the DRL Module box in the DADR pipeline at the top is expanded in the left large box. 

\subsubsection{DRLModule}

The module consists of two main components, a variational autoencoder (VAE) for reconstruction and a generative adversarial network (GAN) for adversarial training. We train the VAE component for in-domain reconstruction, where reconstruction loss is minimized to encourage the encoders and generators to be inverses to each other. The GAN component for cross-domain translation is trained to encourage the disentanglement of the latent space, decomposing it into content and style subspaces \cite{mathieu2016disentangling,narayanaswamy2017learning}. Similar to Huang's work \cite{huang2018multimodal}, the DRLModule consists of several jointly trained encoders $E_{c_1}$, $E_{c_2}$, $E_{s_1}$, $E_{s_2}$, generators $G_1$, $G_2$ and discriminators $D_1$, $D_2$, where $c_i=E_{c_i}(x_i)\sim p(c_i)$ and $s_i=E_{s_i}(x_i)\sim p(s_i)$ for $i=1, 2$. Specifically, the generators are trying to fool the discriminators by successful cross-domain generation with swapped style code. Due to the disentangled style code $s_i\in\mathcal{S}_i$, the underlying mapping is assumed to be many-to-many. We have $p(c_1)=p(c_2)$ upon convergence, which is the shared content space that preserves anatomical information. The overall loss function is defined as the weighted sum of the three components: \begin{equation}
L_{total} = \alpha L_{recon}+\beta L_{adv}+\gamma L_{latent}
\end{equation}

(a) In-domain reconstruction, $L_{recon}=L_{recon}^{1}+L_{recon}^{2}$
\begin{equation}
 L_{recon}^i=\mathbb{E}_{x_i\sim X_i}||G_i(E_{ci}(x_i),E_{si}(x_i))-x_i||_1
\end{equation}

(b) Cross-domain translation, $L_{adv}=L_{adv}^{1\rightarrow2}+L_{adv}^{2\rightarrow1}$
\begin{equation}
L_{adv}^{1\rightarrow2}=\mathbb{E}_{c_1\sim p(c_1), s_2\sim p(s_2)}[log(1-D_2(x_{1\rightarrow2}))]+\mathbb{E}_{x_2\sim{X_2}}[log(D_2(x_{2}))]
\end{equation}

(c) Latent space reconstruction, $L_{latent}=L_{recon}^{c_1}+L_{recon}^{s_1}+L_{recon}^{c_2}+L_{recon}^{s_2}$
\begin{equation}
L_{recon}^{c_1}=||E_{c_2}(x_{1\rightarrow2})-c_1||_1, L_{recon}^{s_2}=||E_{s_2}(x_{1\rightarrow2})-s_2||_1
\end{equation}

\subsubsection{Domain Adaptation with Content-only Images} Once the disentangled representations are learned, content-only images can be reconstructed by using the content code $c_i$ without style code $s_i$. For both CT and MR, their content codes are embedded in a shared latent space that incorporates the anatomical structure information and excludes the modality appearance information. We train a segmentation model on content-only images from CT domain and apply it directly on content-only images from MR domain.

\subsubsection{Joint-domain Learning} 
Joint-domain learning aims to train a single model with data from both domains that works on both domains and outperforms models trained and tested separately on each domain. Our framework can easily generalize to joint-domain learning by including content-only images from both domains for the training segmentation module.

\subsubsection{Implementation details} The SegModule is a standard UNet \cite{ronneberger2015u}.   
Content encoders consist of convolutional layers and residual layers followed by batch normalization, while style encoders consist of convolutional layers, a global average pooling layer, and a fully-connected layer. Generators take the style code (vector of length 8) and content code (feature map of 64x64x256) as inputs. A multilayer perceptron takes the style code and generates affine transformation parameters. Residual blocks in the generator are equipped with an Adaptive Instance Normalization (AdaIN) layer to take affine transformation parameters from the style code. Discriminators are convolutional neural networks for binary classification. For the loss function, $\alpha = 25, \beta = 10, \gamma = 0.1$ in our experiments. Experiments were conducted on two Nvidia 1080ti GPUs. The training time each fold is $\sim5$h for DRLModule and $\sim2$h for SegModule. Testing is very quick.

\section{Experiments and Results}
\subsection{Datasets and Experimental Setup}
We tested our methods on unpaired CT slices of 130 patients from LiTS challenge 2017 \cite{christ2017lits} and multi-phasic MRI slices of 20 local patients with HCC (see Fig.~\ref{fig1}). CT and MR were divided into 5 folds for subject-wise cross-validation. A supervised UNet \cite{ronneberger2015u} trained and tested on pre-contrast MR serves as upper bound of domain adaptation, while a supervised UNet trained on CT and tested on pre-contrast MRI, without domain adaptation, provides the lower bound, which shows the relatively large domain shifts between CT and MRI (see Table~\ref{tab2}).

For our DADR model, in experiment 1, 4 folds of CT and 4 folds of pre-contrast MR were used to train DRLModule, 4 folds of CT were used to train SegModule and 1 fold of pre-contrast MR were used to test SegModule.
In experiment 2, it was the same as experiment 1, but 4 folds of pre-contrast MR were also used to train SegModule. 
In experiment 3, it was the same as experiment 1, except that pre-contrast MR were replaced with multi-phasic MR.

~\\
\noindent
\begin{minipage}{.400\textwidth}
\centering
\includegraphics[width=0.77\textwidth]{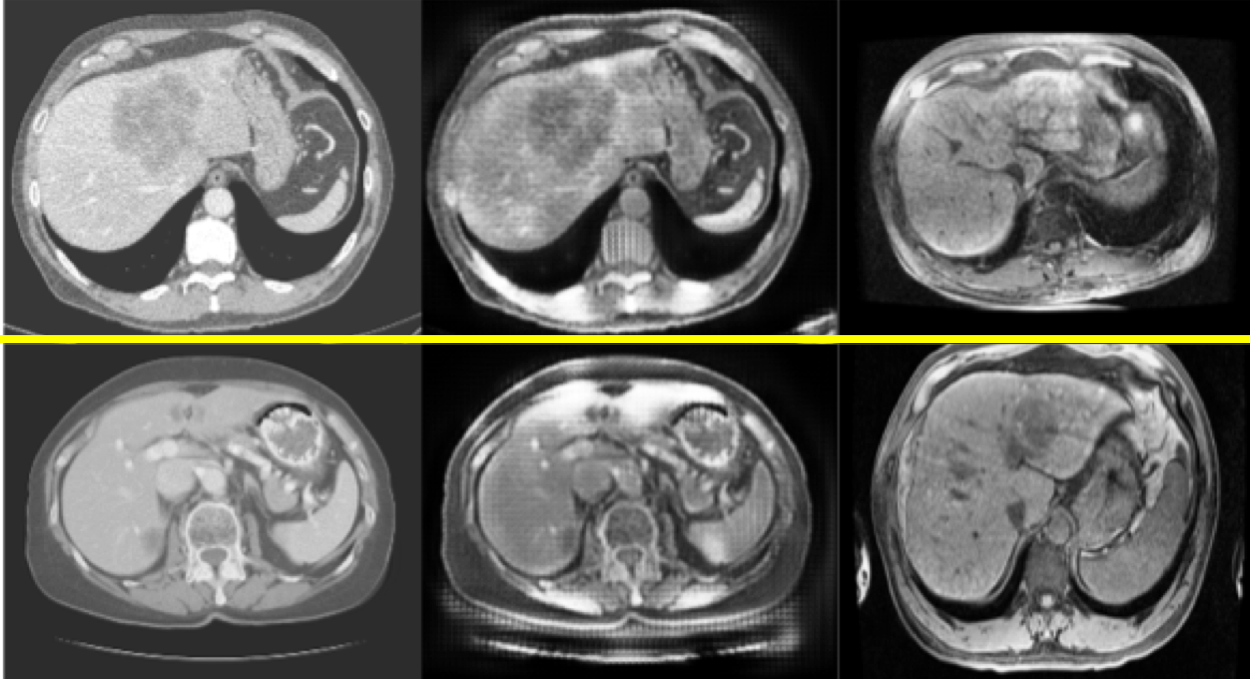}
\captionof{figure}{Two examples of style transfer with CycleGAN, Left to right: CT, generated MR, MR}
\label{fig3}   
\end{minipage}
\hfill%
\begin{minipage}{.58\textwidth}
\centering
\includegraphics[width=0.66\textwidth]{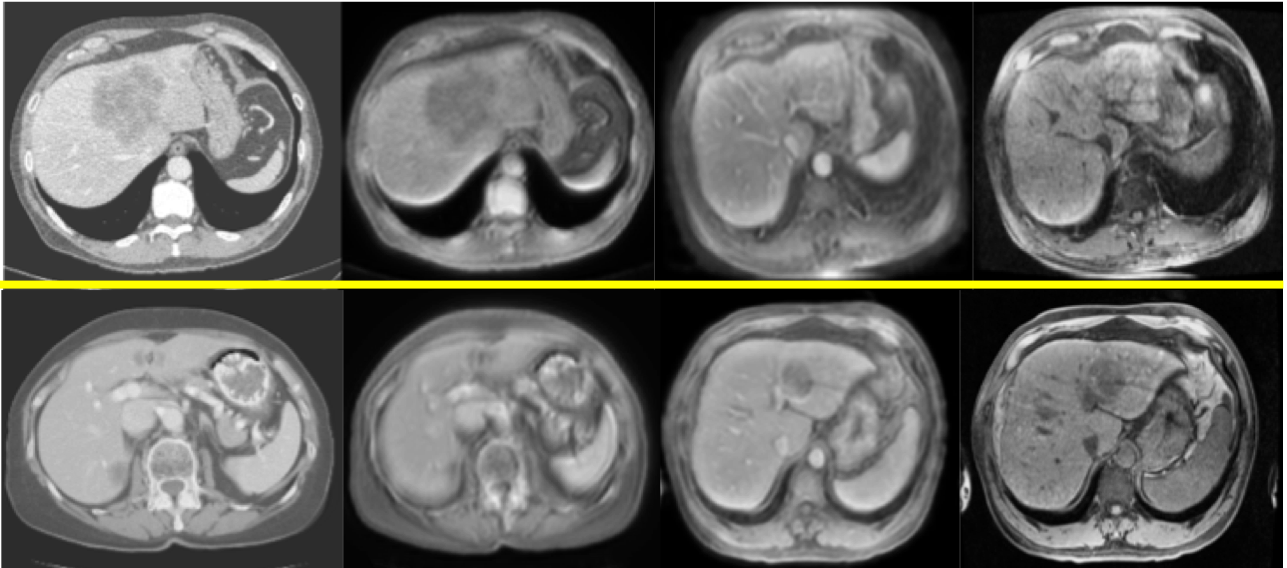}
\captionof{figure}{Two examples of content-only images via disentangled representations, Left to right: CT, content-only CT, content-only MR, MR}
\label{fig4}   
\end{minipage}%

\subsection{Results}
\subsubsection{Experiment 1: Segmentation with Domain Adaptation}
We evaluated Domain Adaptation with CycleGAN (DACGAN) and Domain Adaptation via Disentangled Representations (DADR) respectively and compared with UNet without Domain Adaptation (UNet w/o DA) (see Fig. ~\ref{fig5} and Table~\ref{tab1}).

~\\
\noindent
\begin{minipage}{.3\textwidth}
\centering
\captionof{table}{Comparison of Segmentation Results}
\label{tab1}  
\begin{tabular}{|l|l|l|}
        \hline
        Method & DICE& std\\
        \hline
        UNet w/o DA&  0.26 & 0.07\\
        DACGAN & 0.72& 0.05\\
        {\bfseries DADR}  &  {\bfseries 0.81} & {\bfseries 0.03}\\
        \hline
      \end{tabular}
\end{minipage}%
\hfill%
\begin{minipage}{.6\textwidth}
\centering
\includegraphics[width=0.86\textwidth]{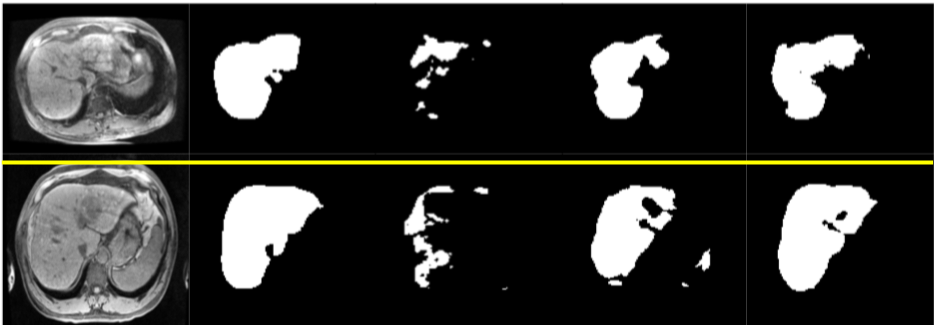}
\captionof{figure}{Two examples of segmentation results. Left to right: pre-contrast MR, liver mask, predictions from UNet w/o DA, DACGAN, DADR}
\label{fig5}            
\end{minipage}
\subsubsection{\textit{DACGAN}}
For comparison purposes, we trained a CycleGAN model with unpaired CT and pre-contrast MR images and performed style transfer on test CT images to generate synthetic MR images. A standard UNet was trained on synthetic MR images and validated on real precontrast MR images. It achieved a DSC score of 0.72 (see Table ~\ref{tab1}) with subject-wise 5-fold cross-validation. Fig. ~\ref{fig3} shows two examples of synthetic MR generated by CycleGAN.

\subsubsection{\textit{DADR}}

Our DRLModule embeds cross-domain images into a shared content space and generates content-only images. Fig. ~\ref{fig4} shows two examples of content-only images via Disentanglement Representations. We trained the UNet model on content-only CT and validated on content-only MR with 5-fold cross-validation and achieved a DSC score of 0.81 (see Table ~\ref{tab1}).

\subsubsection{Experiment 2: Joint-domain Learning}

Besides domain adaptation, our model can achieve joint-domain learning by feeding UNet with both content-only CT and content-only MR as training data.
A single model that works on both CT and MR modality was obtained and outperformed two fully-supervised standard UNet models separately trained on each modality (see Table \ref{tab2}).

\subsubsection{Experiment 3: Multi-modal Target Domain}
We considered multi-phasic MR with three phases as multi-modal target domain with complex underlying distribution and conducted domain adaptation and style transfer on it. 

\subsubsection{\textit{A. Robust Domain Adaptation}}
\begin{figure}[!t]
\begin{center}
\includegraphics[scale=0.34]{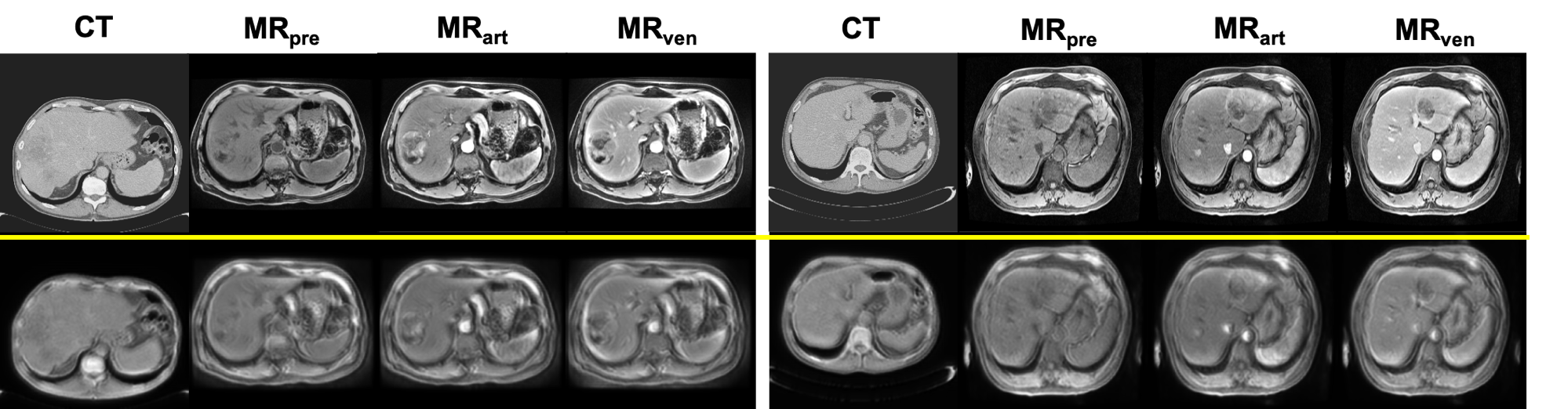}
\end{center}
\caption{Two examples of content-only images via disentangled representations for multi-phasic MR. first row: original images, second row: content-only images. } \label{fig6}
\end{figure} 
A CycleGAN model was not able to handle such large variety in the multi-modal target domain e.g. multi-phasic MR. However, for our model, the shared content space provided a robust representation for anatomical information in multi-phasic MR; thus it remained effective even faced with the multi-modal target domain (see Fig. ~\ref{fig6} and Table ~\ref{tab3}).

~\\
\noindent
\begin{minipage}{.65\textwidth}
\centering
\captionof{table}{Joint-domain Learning with DADR}
\label{tab2}   
\begin{tabular}{|l|l|l|}
\hline
Method & CT tested DSC & MR tested DSC\\
\hline
CT trained&  0.901(0.020) & 0.260(0.072)\\
MR trained& 0.134(0.091)& 0.869(0.044)\\
{\bfseries Joint CT\&MR}  &  {\bfseries 0.912(0.012)} & {\bfseries 0.891(0.040)}\\
\hline
\end{tabular}
\end{minipage}%
\hfill%
\begin{minipage}{.3\textwidth}
\centering
\captionof{table}{DA on Multi-modal Target Domain}
\label{tab3}   
\begin{tabular}{|l|l|l|}
    \hline
    Method &DICE& std\\
    \hline

    DACGAN & 0.52&0.06\\
    {\bfseries DADR} & {\bfseries 0.74} &{\bfseries 0.04}\\
    \hline
\end{tabular} 

\end{minipage}

\subsubsection{\textit{B. Diverse Style Transfer}}

\begin{figure}[!t]
\begin{center}
\includegraphics[scale=0.34]{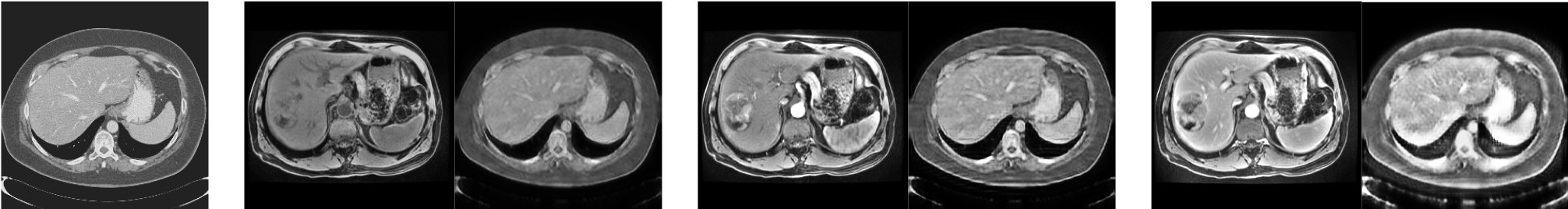}
\end{center}
\caption{Three examples of multi-modal style generation with reference, from left to right: CT, three pairs of reference MR and generated MR (pre-contrast, arterial, portal venous phase)} \label{fig7}
\end{figure}

Furthermore,  our model can realize diverse style transfer by changing the style code while preserving the anatomical structure with the same content code. The style code can be randomly sampled in the style space or encoded from a reference image by style encoder $E_{s_i}$ (see Fig. ~\ref{fig7}).

\subsection{Analysis}
We tested our model on unpaired CT and MRI data. It is noteworthy that it is a highly unbalanced cross-domain data, where CT is of better quality and the size of the CT dataset is about 6.5 times the size of MR.
Experiment 1 shows that our model is superior to CycleGAN in terms of DSC score for cross-modality segmentation with domain adaptation. Experiment 2 shows a promising application of our model for joint-domain learning, which makes learning from unpaired medical images with different modalities a reality. Experiment 3 shows robustness of our model under multi-modal target domain with large diversity and the potential for diverse multi-modal style transfer. The disentangled representation is the key to fulfill the many-to-many mapping assumption and recover the complex relationship between two domains. It is also of vital importance to discover the shared latent space that preserves the semantic feature-level information.
\section{Conclusions and Discussions}

We proposed a cross-modality domain adaptation pipeline via disentangled representations, which may improve current clinical workflows and allow for robust intergration of multi-parametric MRI and CT data. Instead of one-to-one mapping, our model considers the complex mapping between CT and MR as many-to-many and preserves semantic feature-level information, thus ensuring robust cross-modality domain adaptation for a segmentation task. We validated and compared our model on CT and pre-phase MR from HCC patients with state-of-the-art methods. With multi-phasic MRI, our model demonstrated strong ability to handle multi-modal target domains with large variety. Furthermore, our model had good generalization towards joint-domain learning and showed the potential for multi-modal image generation, which can be further investigated in the future. In addition, we could focus on specific anatomical structures in the content space, such as liver and tumor, by including task-relevant loss to improve the results further. 

\bibliographystyle{splncs04}
\bibliography{reference}
%





\end{document}